**Title**: Do software firms collaborate or compete? A model of coopetition in community-initiated OSS projects

**Abstract**. [Background] An increasing number of commercial firms are participating in Open Source Software (OSS) projects to reduce their development cost and increase technical innovativeness. When collaborating with other firms whose sought values are conflicts of interests, firms may behave uncooperatively leading to harmful impacts on the common goal. [Aim] This study explores how software firms both collaborate and compete in OSS projects. [Method] We adopted a mixed research method on three OSS projects. [Result] We found that commercial firms participating in community-initiated OSS projects collaborate in various ways across the organizational boundaries. While most of firms contribute little, a small number of firms that are very active and account for large proportions of contributions. We proposed a conceptual model to explain for coopetition among software firms in OSS projects. The model shows two aspects of coopetition can be managed at the same time based on firm gatekeepers. [Conclusion] Firms need to operationalize their coopetition strategies to maximize value gained from participating in OSS projects.

# 1. Introduction

Increasingly, software products are no longer developed solely in-house, but in a distributed setting, where developers collaborate with "distributed collaborators" beyond their firms' boundary [1, 12]. This phenomenon includes open source software (OSS) communities, crowd-sourcing, and software ecosystems (SECO). This differs from traditional outsourcing techniques in that initiating actors do not necessarily own the software developed by contributing actors and do not hire the contributing actors. Community-initiated OSS projects are an example of the context in which actors coexist and coevolve.

From firms' perspective, it is beneficial for the development of software products whose scopes exceeds their own capabilities by leveraging external resources, exploring opportunities to enter new markets [14], performing an inside-out process [2], and employing strategic recruitments [15]. From communities' perspective, the participation in such environment probably causes firms to open up its successful products and product lines for functional extensions by external developers [1]. Instead of being exclusive and localizing product development, firms are exploring different ways to invite contributions from external actors without revealing core technology, business value and customer relationships [30].

Before the full potential advantages of open sourcing are leveraged, commercial firms need to consider several concerns. At the organizational level, the firm's benefit and the community goals are not always the same [3]. Participation of commercial firms in OSS projects with their diverse motivations and business strategies might introduce variance, and sometimes conflicts in project evolution [14]. Existing research on OSS highlights the role of collaboration with extensive research on communication and coordination practices, patterns and lessons learnt from OSS communities [4-7]. However, there seems to be far less research concerns about the conflicts among firms regarding to their strategic development. Firms attempt to gain competitive advantages from their participation in OSS projects [32]. When there occur mismatches in term of interests and objectives, firms may behave uncooperatively in order to prevent others from achieving their goals [33]. The conflict occurs not only at the managerial level, such as project governance [31], but also at the operational level, such as code contribution, bug fixes, and requirement elicitation [10,11,14,29].

Coopetition, as a business phenomenon, is about collaborating and handling a firm's competitive advantages when participating in OSS projects [8,9]. In a coopetitive environment, firms cooperate with each other to reach a higher value creation compared to the value created without the interaction. The basic assumption for coopetitive relationships is that all activities should aim at the establishment of a beneficial partnership with other firms, including partners who may be considered as a kind of competitor [34]. Since coopetition applies to inter-firm relationships, OSS project offers an ideal context for understanding the phenomenon among firms that develop and utilize a common software codebase [11].

Empirical research on coopetition is scarce, especially studies in Software Engineering (SE) and at the organizational level [33]. Research in this area is probably hidden by the inconsistent treatment of the cross-disciplinary natures of cooperation and competition, and their related constructs. Our research objective is to explore how firms interact and manage the phenomenon of coopetition in OSS projects. To best of our knowledge, there exists only a few studies that examine the phenomenon of coopetition among commercial firms in OSS projects [10, 14, 29, 33]. Research questions (RQs) were derived from this research objective. Firstly, we aimed at understanding the basic foundation on firm participation in OSS projects. Based on this knowledge, we explored further theoretical elements of coopetition. We use here the word "coopetitively" as an adverb of coopetition:

*RQ1: How do commercial firms participate in community-initiated OSS projects?*

*RQ2: How do commercial firms manage coopetition with other firms in such context?*

Our contributions are two folds, firstly we portrayed the situations where both competition and collaboration occurs in OSS projects. Considering the body of knowledge about firm participation in OSS projects, our work confirms some patterns and also extends them by exploring the firm awareness, coopetition and their antecedent factors. Adopting a mixed-method research, we quantitatively examine organizational interaction patterns and qualitatively explore how firms perceive and employ coopetition strategies. Secondly, we theorize constructs of coopetition by proposing a Coopetition in Open Source Software (COSS) model. Previous studies that mention the term "*coopetition*" [10, 14], do not investigate the constructs under this phenomenon. Hence, to our best knowledge, this is among the first studies in SE investigating this concept. The proposed model reveals building blocks of coopetition in OSS firms network and its relationship to consequent factors.

The study is organized as follows: Section 2 presents a background about coopetition and firm participation in OSS projects. Section 3 describes our research methodology, Section 4 presents our findings, and Section 5 discusses the findings. Finally, Section 6 concludes the paper.

# 2. Background and Related Work

## 2.1. The phenomenon of Coopetition

The origin of coopetition is from business research when investigating buyer–seller relationships within a business network [8,9]. The trade-off between cooperation and competition is emphasized as a mean of creating a progress among actors involved in long-term relationships. Coopetition conceptualizes the interaction among firms in relation to their strategic development [8, 9]. Dagnino et al. defined coopetition as "*a kind of inter-firm strategy which consents the competing firms involved to manage a partially convergent interest and goal structure and to create value by means of coopetitive advantage*" [26]. The authors proposed two forms of coopetition, a dyadic coopetition (concerns among two-firm relationships) and a network coopetition (involving more than two firms, i.e. value chain) [26]. Bengtsson argued that a dyadic relationship is a paradox that emerges when two firms cooperate in some activities, such as in a strategic alliance, and at the same time compete with each other in other activities [9]. It means that actors within a firm need to be divided to take charge of either collaboration or competition.

Coopetition can occur in a more complex form, with a network of firms. The coopetition strategy can be applied at a micro level (among functional and divisional departments in a firm), a meso level (among firms in the same industry, between vendor and supplier) and a macro level (among cluster of firms or firms across industries) [26]. Literature also discusses some antecedent factors relating to coopetition at the micro level, such as shared vision, perceived trust and perceived benefits [35]. A study points out some possible impacts of coopetition on knowledge sharing and job/ task effectiveness [35]. By selecting a highly innovative OSS project that contributes to firms' strategic values, we illustrate dependencies between competitors due to structural conditions, why and how competitors cooperate

## 2.2. Collaboration in OSS projects

Collaboration is an aspect of coopetition that is much explored in OSS projects. It is common to look at OSS projects' archives to reveal communication, collaboration and coordination approaches, frequency, patterns and best practices at different level of analysis [12,19,44-51]. Early research has observed an onion-like structure of contribution in OSS projects [44-47]. At the center of the onion are the core developers, who contribute most of the code and take care of the design and evolution of the project. In the next ring out are the co-developers who submit patches (e.g. bug fixes), which are reviewed and checked in by core developers [48]. Further out are the active users who do not contribute code but provide use-cases and bug-reports as well as testing new releases. The awareness of people and activities through OSS social structures enhances collaboration effectiveness and ensures that little effort is wasted in duplicate work [50]. A large amount of studies investigates the combination of social and technical aspects of OSS projects, by analyzing a social network created by contributors who work and communicate in the same set of files [52-55]. Bird et al. showed that a socio-technical network of software modules and developers is able to predict software failure proneness with greater accuracy than other prediction methods [54]. Wolf et al. formed a developer-task network to explore the impact of developer communication on software build integration fail [55]. A common assumption of these studies is that developers behave regardless of their commercial affiliations in OSS projects, indicating by unweighted analysis approaches when formulating the social networks. In case a significant number of developers from firms contributes to the project, organizational features, such as firms' strategies and governance mechanism might influence the communication structures of the OSS projects. In this work, we will use the social network analysis (SNA) to investigate interaction patterns, i.e. collaboration and competition in OSS projects. While we also form the developer-task-developer network, the difference is that the relationship is analyzed at firm level.

## 2.2. Firm participation in OSS projects

Firm participation in OSS communities has been studied from different angles, leading to different observations. From the firm's perspective, social structures as those in OSS enables the integration of external resources [56]. If a software firm can attract OSS contributors, its development costs might decrease, as they don't have to pay for these voluntary contributions. Moreover, firms can extend their distribution channels and innovate their business models with the evolution of OSS projects. The participation of commercial firms implies that community work has somehow to be aligned with what happens inside the boundaries of the firm. As community and firm interests do not often entirely overlap [57], firms seek options to influence community work to avoid such situations. Dewan et al. showed that the heterogeneity, which exists between firm-paid developers and voluntary developers shapes the evolution of a OSS community and its product [58]. Dahlander et al. studied the network of relationships within the GNOME project, discovering that the presence of hired developers often generate an initial diffidence among unpaid programmers [25]. Comino et al. found that recent entrance of firms is likely to change the forces driving the evolution of OSS projects [59]. As Gallivan noticed, strong explicit governance approaches (i.e., rules and norms provided in the documentation and agreements) can directly affect other firm's benefits [64]. In community-based OSS projects that involve multiple commercial firms, the influence of firms to the projects might introduce conflicts and even competition among the firms. This observation implies that coopetition among firms might exist in OSS context, but do not reveal how the phenomenon occurs and operate particularly.

Research investigating the phenomenon of coopetition in SE is scarce. Teixeira et al. found that competition for the same revenue model (i.e., operating conflicting business models) does not necessarily affect collaboration within OSS projects [29]. Valenca et al. explored the concepts of competition and collaboration in requirement engineering processes [14]. The authors concluded that even though competition was inevitable among companies, establishing a long-term partnership was a crucial driver for innovation and performance. Linaker et al. investigated stakeholders' influences and collaboration patterns in Hadoop project [10]. The authors showed that regardless of business models, all firms work together towards the common goal of advancing the shared platform [10]. Our work extends this knowledge by a comprehensive conceptualization and empirical investigation of coopetition. From both qualitative and quantitative data, we were able to derive a model capturing the coopetition among OSS participating firms.

## 2.3. A theoretical model of Coopetition in OSS projects

A theoretical model links theoretical elements in a certain semantic manner, i.e. a causal relationship, helping to design data collection and analysis. Literature reveals factors that lead to the occurrence of collaboration and competition (antecedent factors), and their impact on firms' outcomes (consequent factors). It is noted that we do not aim for model completeness, but for a foundation of further investigation. The further investigation would discover which factors valid in the context of software industry, particularly OSS projects.

As seen in Figure 1, coopetition is the studied construct, and it is linked to its antecedent factors, i.e. structural condition, strategic vision, trust and perceived benefits [35, 36, 37, 38, 40, 41, 42]. **Strategic vision**: sharing strategic vision is essential for cooperation at team level [35], as the vision reflects important agreements of beliefs and assumptions that consequently bring internal stability to the cooperative attitude [36]. At the strategic level, vision typically is about the firm's value and business development. Shared vision draws a roadmap for the organization or firm, setting the priorities for their team planning and implying its critical determinant role in lessening malign competition [35]. The vision can be shared via meetings or workshop with high-level managers.

**Trust**: is considered as a relationship of reliance among members of a team or an organization. Trust is defined as "*the willingness of a party to be vulnerable to the actions of another party based on the expectation that the other will perform a particular action important to the trustor, irrespective of the ability to monitor or control that other party*" [39]. The importance of trust in the success of interpersonal relationships is reported previously in OSS projects [37,38]. Moreover, trust is the key of transforming OSS as a community of individual developers, to OSS as a community of firms [40]. The cooperation that captures the level of coordinated actions between team members in their efforts to achieve mutual goals cannot be realized without trust among the members.

**Perceived benefit**: on one hand, perceived benefits are associated with a cooperative attitude, involving compatible interests as common benefits can motivate collaboration, leverage team or person's capabilities for obtaining such benefit [41]. In OSS projects, perceived benefits of participating in the communities are reduced development cost, community knowledge, and reduced maintenance cost. On the other hand, perceived benefit is also associated with a competitive attitude. Individuals are likely to pursue their own objective at the expense over all team's goal [42]. This could be applicable for organization in an ecosystems or supply chains. The more benefit a firm perceive for obtaining a conflicting artifact or resource, the more they likely to compete over the resource [35].

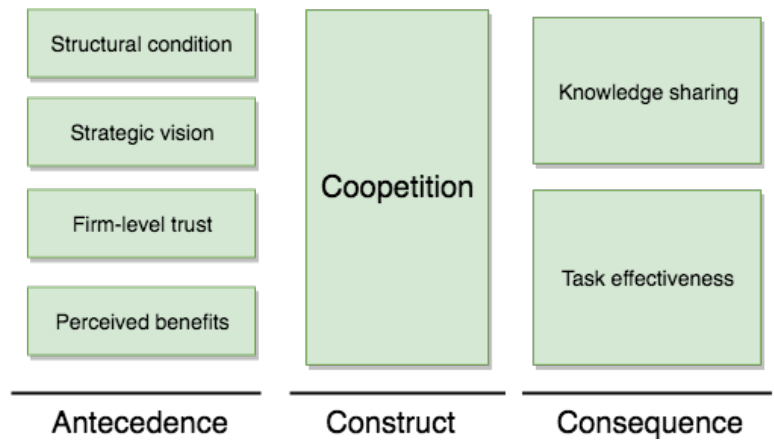

**Figure 1: A theoretical framework of Coopetition (adapted from [35])**

In our theoretical framework, coopetition is also associated to its consequent factors, i.e. knowledge sharing and task effectiveness [35,43]. **Knowledge sharing** at organizational levels is seen as sharing of organizational experience and knowledge, i.e. technical know-how, domain expertise, work practice, etc with other collaborators, and hence increasing the overall knowledge in the joint project [35]. As knowledge is a critical source of competitiveness, managing knowledge sharing among members of an organization plays a prominent role in sustainable competitive advantage [43]. **Task effectiveness** in team collaboration represents individuals' perceived capacity of conducting collaborative tasks, whereas knowledge sharing enhances the ability of collaborator's knowledge exchange.

## 3. Research Approach

### 3.1. Study design

We conducted this work by using a two-phase multiple-case study design [27]. The phases in the research occur due to the discrete continuation of our internal research project. Compared to descriptive and confirmative case studies, exploratory case studies are suitable for the first phase research as we would like to discover the phenomenon of coopetition, whether it exists, in which form and its relationship to its context setting. This phase was done as a part of a master thesis. In the second phase, we conducted a descriptive study on describing collaboration, competition in the selected cases. In the third phase, we found another case study to confirm the qualitative findings. This step was conducted to validate what we observed in the first two cases. We followed the guideline by Runeson and Höst [66] to execute case study, including case selection, data collection and analysis.

Case selection is not straightforward. There are abundant OSS projects available; many of them are abandoned or individual efforts. A brainstorm session was conducted among the paper's authors to decide case selection criteria as below:

- Commercial participation: the OSS project should have multiple commercial firms participating in the development. In addition, there must be an adequate way to identify them.
- Successful and on-going: the OSS project must be successful and on-going. This implies that the project attracts developers and the development of the software is progressing.
- Active projects with many activities: the OSS project must have a high level of communication and code commits in the project, showing by rich data archive.

By reviewing literature on OSS projects in SE, we learnt several OSS projects that were commonly investigated in SE research, such as Apache, Mozilla, Eclipse and Linux [71]. The selected cases should not only satisfy the selection criteria, but also novel in SE research. We were suggested to Wireshark by a colleague who participated in the project. Many reasons contributed to this choice. Firstly, the contributor list and community activity revealed high participation and involvement of commercial companies. Wireshark is a typical instance of a OSS project. The project uses software informalisms for development collaboration, the developers are a mix of firm-paid developers and volunteers, and the software is licensed under the GNU General Public License (GNU GPL). Wireshark is also a very successful on-going OSS project, with a high number of contributors and active users, consistently pushing development forward. Having selected Wireshark as the first case, we proceeded to find and select the second case for our study. To be able to do a literal replication, the second case should have similar properties as the first case. After a long period of searching, we ended up with three promising cases that matched the specifications: Horde, Samba and Wine. From the comparison it was evident that Samba was very similar to Wireshark, i.e. both projects were licensed under GNU GPL, both projects had many firms participating, and they both had a yearly conference where developers cane together to discuss further development and socialize. We planned to have the third case to validate the qualitative findings from Wireshark and Samba. Among several OSS projects we attempted to contact, Bootstrap developers were the one agreed to participate in the study.

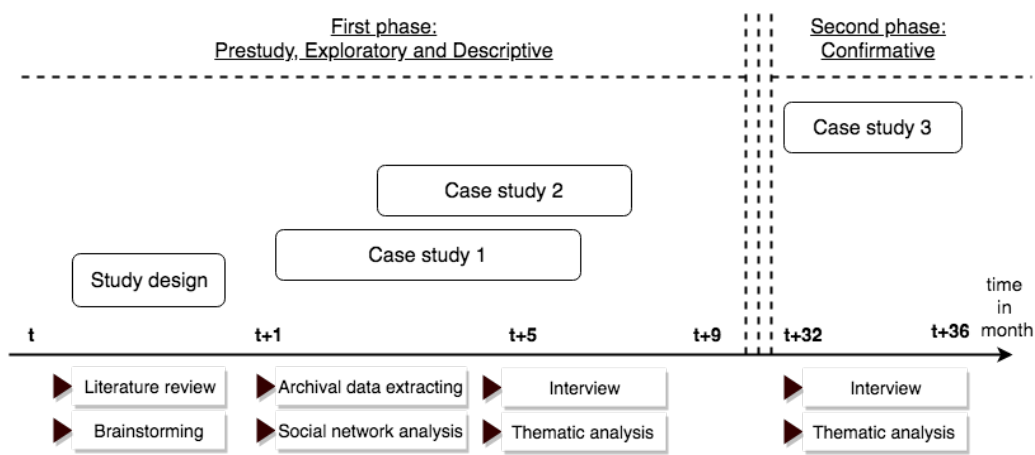

**Figure 2: Overview of the research process**

The research process is described in **Error! Reference source not found.**. At the pre-study phase, literature review and brainstorming with experts were done to come up with research objective and study design. At the exploratory and descriptive phase, the first two cases were investigated for understanding how commercial firms participated in OSS projects, if the phenomenon coopetition exists and in which form. As the explorative nature of this phase, a wide range of topics was discovered, such as collaboration patterns, firm awareness, competition, code practices, etc. The data were extracted from project archive, i.e mailing lists, bug tracking system and code repository. In this phase, we also collected qualitative data, i.e. interviewing relevant stakeholders to explore in-depth phenomenon observed from the quantitative data. At the confirmative phase, we conducted some interviews to confirm and to validate the observation from the first two cases.

## 3.2. Case description

Wireshark[1] is an OSS toolkit developed by a community of networking experts around the world under the GNU General Public License. The project is officially operated under the Wireshark name since May 2006. Out of the 802 developers listed in Wireshark contributor list, 342 were classified as firm-paid developers (43%). The remaining 460 developers (57%) were classified as volunteering developers. The firm-paid contributions come from 228 firms.

Samba[2] is an OSS suite that provides file, print and authentication services to all clients using the SMB/CIFS protocol. Samba is licensed under the GNU General Public License, and the Samba project is a member of the Software Freedom Conservancy. In Samba, 316 developers were evaluated, where 182 (57%) of them were classified as firm-paid developers. The contributions come from 45 firms. Communication and collaboration between developers in the Wireshark and Samba community mainly occur in two places; the developer mailing list and the bug tracking system.

Later, a third OSS project was selected as a more recent project to provide complementary qualitative data. Bootstrap[3] is a frontend Javascript-based framework for developing responsive, mobile first projects on the web. The project was released as an OSS project since 2011 under MIT license. Bootstrap were contributed by large firms, such as Twitter and Github. At the time the research was conducted, Bootstrap has been the most-starred project on Github, with over 90.000 stars and more than 38.000 forks. The communication in Bootstrap was done via many channels, i.e. StackOverflow, Slack, and Github tracker. Source code and issue management was done via Github.

---

[1] https://www.wireshark.org

[2] https://www.samba.org

[3] http://getbootstrap.com

## 3.3. Data collection

Both quantitative and qualitative data was collected and described in Section 3.3.1 and Section 3.3.3 correspondingly. We present our approach in identifying a firm in data from projects' archives in Section 3.3.2.

### 3.3.1. Quantitative data

The main source of quantitative data is from mailing lists, code and issue repositories, as they are common data sources when studying OSS [4, 10, 19, 22]. We collected three types of data, namely developer profile, firm profile and communication data. The developer profile was found from project public pages, such as project wiki and confluence page. Basic information, like developers' email addresses and timestamp of file commits were extracted from JIRA and GIT. From developers' profiles, we were also able to identify the list of firms in a OSS project. An invitation for interview was sent in a snowballing manner. After firm-paid developers accepted our invitation for interview, basic information about the firm was required by us. Besides, firm information was also collected from online sources, such as company website, and published materials. The communication data was collected from two main sources, namely issue tracking system and mailing list. These sources contained detailed information about events and activities that had occurred in the communities several years back in time. Table 1 gives an overview of when the sources were first used and how many entries they have today in Wireshark and Samba.

**Table 1: Summary of quantitative data from Wireshark and Samba**

| Project | Data source | Date of first entry | # of entries |
|---------|-------------|---------------------|--------------|
| Wireshark | Mailing list | 31.05.2006 | 27230 |
| | Bug tracking system | 08.04.2005 | 7862 |
| | Code repository | 16.09.1998 | 42794 |
| Samba | Mailing list | 03.01.1997 | 90588 |
| | Bug tracking system | 24.04.2003 | 9659 |
| | Code repository | 04.05.1996 | 84699 |

### 3.3.2. Identification of firm participation

Information whether a participant is a firm-paid or volunteer developer, is not generally available in OSS projects. Consequently, we needed to come up with a classification technique to identify firms' participation. The approach has been successfully used in a previous study [63]. The following information was evaluated in the process of classifying the developers:

- Current status in the community: active or not any more

- Email domain: The email domain used by a developer can reveal firm association. We regard it as unlikely that a developer use a job email to participate in an OSS project if it is not related to the job as a paid developer. This measure is the most distinctive classification entity.
- Email signature: Some developers have their employment firm name as part of their email signature, which they use when posting to the mailing list or bug tracker.
- Personal homepage: Searching for a developer's name on the web can give directions to a personal homepage or blog that might reveal company association.
- Social networks: Searching for a developer's name on social networks like LinkedIn and other professional pages might reveal firm affiliation.
- Presentations and conferences: Developers that give presentations commonly include name and firm in the presentation slides, which are easy to find by a web search.

Some issues were faced when identifying contributors' affiliations. Firstly, there is a different level of contributions in OSS projects. There is often a lack of information about what is required to become a contributor. Moreover, majority of the participants in the mailing list only posted one mail, which makes it a waste of time and effort to identify these participants as the contribution towards the firm's interaction and software development is minuscule. We decided to exclude developers with less than ten entries in the mailing list or bug tracking system. Secondly, matching name, alias and email address is not always straightforward. In Wireshark, the spam protection policy hides the full email address, for instance: "*From: [developer name] <name@x>*". Moreover, entries in the bug tracking system have email listed, but no name. The code repository entries in Wireshark does not contain name or mail of the developer, instead a username or a nickname is used. We had to use project wiki pages and personal contacts with some core developers of the project to provide mapping of most of the usernames to the actual developers.

### 3.3.3. Qualitative data

Regarding to qualitative data, interviews were selected from a convenient sample consisting of the firm-paid developers from Wireshark, Samba and Bootstrap. Ten interviews were conducted as seen from Table 2. In Wireshark and Samba, we managed to have interviews from firms in a core layer and a peripheral layer (detail as shonw in Figure 6). Due to non-disclosure agreements, we did not reveal the actual identity of companies (quantitative data was publicly available, hence did not have this constraint). We used alias D1 to D10 to represent for such firms.

As we did not know much about the population, we aimed for a non-probabilistic sampling technique using a conjunction of purposive and snowball sampling. In Wireshark, we used an existing connection to one of the core contributors as a starting point, and asked for suggestion of developers that could be interesting to interview next. The core contributor pointed out relevant developers for the research topic, and assisted in contacting them by posting our interview invitation on the core contributor mailing list. In Samba, we selected relevant developers in the OSS project based on the quantitative data and sent interview invitations to these by email. In Bootstrap, we had a developer actively contributing to the project in our personal network. From him, we got two more interviews with firm-paid participants in Bootstrap.

The interview guide consisted of both closed and open questions. The closed questions were mainly used in the introduction phase of the interview to solicit background information about the respondent, firm and OSS project context. In addition, the closed questions were used to confirm or attribute statements given by other developers. The open questions were used to collect information about: (1) work process/bridge engineer role, (2) firm awareness/organizational boundary and (3) position in the community/contributions. The interview guide and interview questions is publicly available[4]. The interviews were conducted in English, except for one in Norwegian. The duration of the interviews ranged from 45 minutes to 72 minutes. All the interviews were recorded to facilitate subsequent analysis and minimize potential data loss due to note-taking. These recordings were thereafter transcribed verbatim. Transcribing audio records resulted in 55 pages of rich text.

**Table 2: Summary of interview profiles**

| Alias | Domain | Firm type | Firm size | OSSs |
|---|---|---|---|---|
| **D1** | Telecommunication | Corp. | 10 000+ | Wireshark |
| **D2** | Wireless networking services | SME | 11 - 50 | Wireshark |
| **D3** | Messaging system | SME | 11 – 50 | Wireshark |
| **D4** | Telecommunication | Corp. | 10 000+ | Wireshark |
| **D5** | IT security services | Corp. | 51 - 200 | Samba |
| **D6** | Server and OS development | Corp. | 10 000+ | Samba |
| **D7** | Telecommunication | Corp. | 10 000+ | Samba |
| **D8** | Social media | Startup | 1 - 10 | Bootstrap |
| **D9** | Hosting and file sharing | SME | 51-200 | Bootstrap |
| **D10** | Social media | Startup | 1 - 10 | Bootstrap |

## 3.4. Data analysis

### 3.4.1. Social network analysis (SNA)

SNA is a common approach to investigate communication and collaboration patterns based on data from mailing lists or issue tracking systems [52-55, 65]. This has been extensively used for constructing a developer-task network and measuring different network features [52-55].

---

[4] https://tinyurl.com/OSScoopetition-InterviewGuide

We adopted this approach in firm level, to understand the collaboration pattern among firms via communication networks. Consequently, we used the firms as nodes and the interaction between firms as edges. Interaction among firms is represented by communication via either a mailing list or comments on an issue tracking system. The SNA was done in four steps:

- **Step 1:** Construct discussion trees from a mailing list and an issue tracking system. A discussion tree consists of an identifier node, a source node and a set of responder nodes (which can range from none to many). The developer that initiates a discussion is regarded as the source, and the developers that follow-up on a discussion is regarded as responders

- **Step 2**: Filter the discussion trees to remove messages with noises (irrelevant information). As shown in Figure 3, we convert a discussion tree to an undirected graph.

- **Step 3:** Give firm's affiliation to nodes in the graph, so that the interaction could be grouped at a firm level, rather than at individual level.

- **Step 4:** Build the social network by using NodeXL[5] tool.

We were interested in the position of a firm within the context of the entire network, leading to the adoption of metrics, i.e. degree centrality, betweenness and closeness [65]:

- Degree of centrality is a measure of the number of links incident upon a firm, i.e. how many other firms that a firm is connected to.

- Betweenness centrality is a measure of the number of a shortest path between two firms that a firm lies on, quantifying the degree to which an individual in a network mediates information flow.

- Closeness centrality measures the distance from a firm to all other firms in the network. Lower values indicate that the component is farther away from all other nodes.

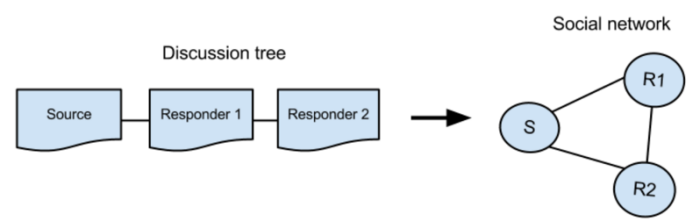

**Figure 3: Constructing SNA from a discussion tree**

---

### 3.4.2. Qualitative analysis:

The analysis of the qualitative data was undertaken following a guideline for thematic synthesis [16]. Thematic analysis allows main themes in the text to be systematically summarized and is also familiar by the first two authors of the paper. The process of how quantitative data from Section 3.4.1 facilitates the qualitative analysis and the use of the theoretical model to guide the analysis is shown in Figure 4. The interviews were prepared for analysis by manual transcription of the audio recordings to text documents, and the email responses were refined to transcripts of the same disposition. This resulted in 55 pages of rich text. Segments of text about firms' interaction, i.e. activities, attitudes about communication, collaboration and competition were identified and labeled. Data from the Bootstrap case showed a level of data saturation, as there was not much new information from the case. After two rounds of reviews of the data, we ended up with 84 codes. The following step of the thematic analysis was to merge the codes and the corresponding text segments into themes. A theme in this context is essentially a code in itself, however, a theme is an increased distanciation from the text, and thus an increased level of abstraction. There are two scenarios with a theme, the first one is that identified text relates to an element in our theoretical model (as in Figure 1). The red arrow in Figure 4 describes such scenario. The second scenario is the theme could be interpreted as a new concept. The green arrow in Figure 4 describes such scenario. By grounded from existing elements and new ones, we are able to come up with an empirical model describing the concept of coopetition in three OSS projects (Section 5).

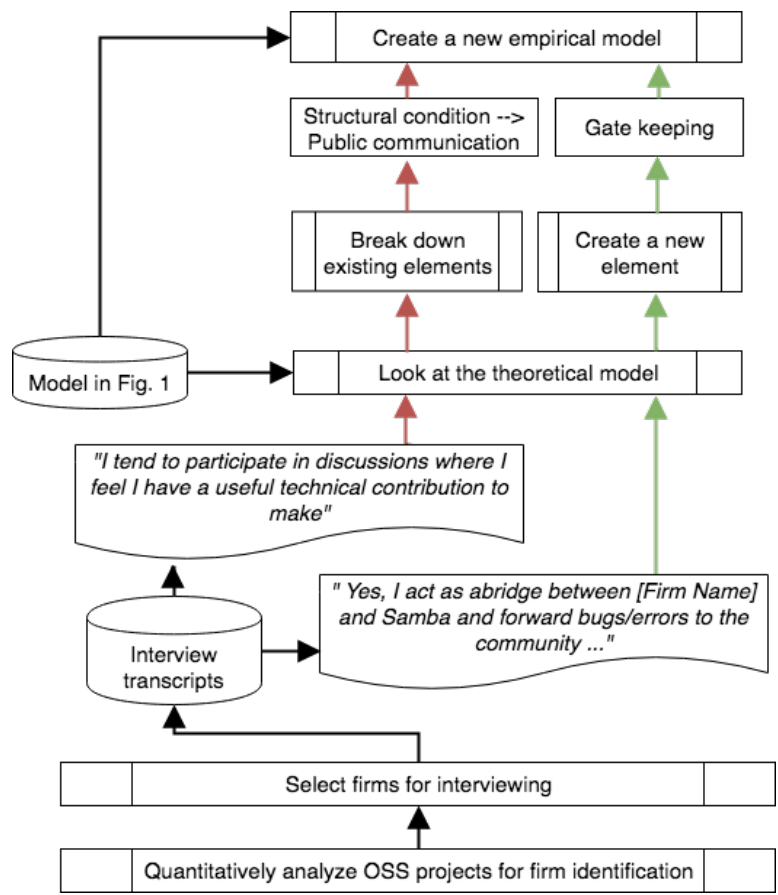

**Figure 4: Steps of qualitative analysis and examples**

# 4. RQ1. How do commercial firms participate in community-initiated OSS projects?

In Section 4 we present the results of the collaboration pattern analysis. Two elements from each OSS project are presented: (1) significance of firms' contribution to OSS projects (Section 4.1), and (2) the social network structure of firms (Section 4.2).

## 4.1. The significance of firm's contribution

Regards to Wireshark project, from the 342 firm-paid developers, 228 unique commercial firms were identified, constituting 43% of total number of contributors. There are only 8% of the firms having three or more developers participating in the community. Firms with the largest number of participating developers are Cisco, Ericsson and Siemen. Whereas, 78 % of the firms have only one developer participating. The code repository log contained 21927 entries, where 12053 of them were committed by firm-paid developers.

Regards to Samba project, there are 182 firm-paid developers representing 90 different commercial firms, constituting of 58% of total number of contributors. In comparison to Wireshark project, Samba is more dominated by firms' contributions. Nine percent of total number of firms have three or more developers participating in the community, and 84% of the firms has only one developer participating. The top ten firms participating in the community with regard to number of developers is presented in Table 3.

**Table 3: Top firms with highest number of developers**

| Wireshark | | Samba | |
|---|---|---|---|
| **Firm** | **# of developers** | **Firm** | **# of developers** |
| Cisco | 16 | IBM | 17 |
| Ericsson | 11 | RedHat | 14 |
| Siemens | 8 | SerNet | 8 |
| Netapp | 6 | SUSE | 8 |
| Citrix | 5 | EMC | 4 |
| Lucent | 5 | SGI | 4 |
| MXTelecom | 5 | Exanet | 3 |
| Nokia | 5 | HP | 3 |
| Axis | 4 | Cisco | 3 |
| Harman | 4 | Canonical | 2 |

## 4.2. The social network structure of firms

We illustrate the constructed SNA based on data from issue tracking systems in Wireshark, as shown in Figure 5. The node represents for a firm and the link between nodes represents for a communication link between them. The node degree was counted, including both in-degree (number of interaction received) and out-degree (number of sent interaction). By looking at the social network of Wireshark, a firm can belong to one of three contribution layers: (1) a core layer with high centrality degree, representing firms that actively communicate with others (for instance, Thales and Ericsson), (2) a peripheral layer with moderate centrality degree, representing firms with a medium number of messages to other firms (for instance, Tieto and Novell) and (3) a passive layer with low centrality degree, representing firms with small amount of message sending in and out (for instance, Broadcom and Motorola).

The contribution from commercial firms in the issue tracking systems conforms to the same pattern as in the mailing list; significant, but highly diversified. In total, the issue activity by commercial firms constitute 39 % in Wireshark and 66 % in Samba. Figure 5 reveals that a small number of firms stay in the core layer and most of the firms locate in the passive layer. The similar network structure was observed in case of Samba project. We do not present the SNA figure for Samba due to limited space.

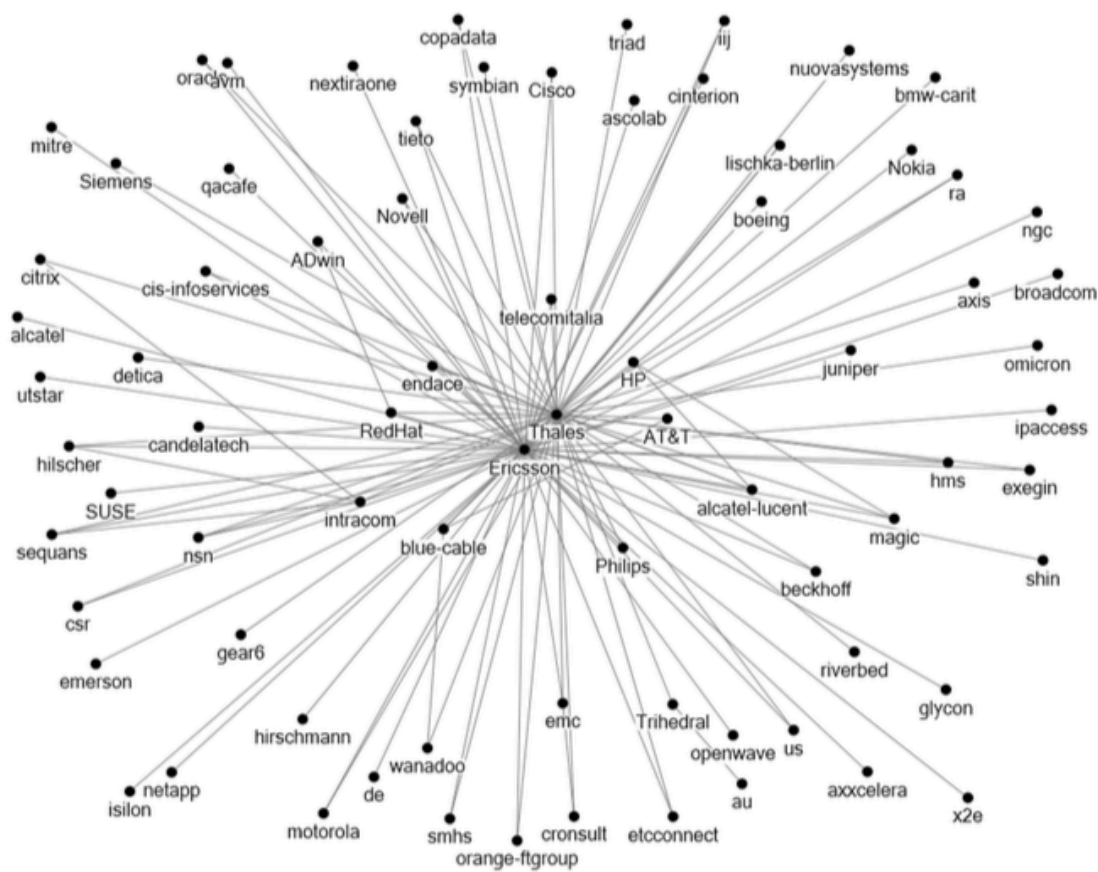

**Figure 5: The social network of Wireshark via issue tracking system**

The collection of identified commercial firms constitutes a large fraction of the activity in the mailing list in both projects, approximately 27 % in Wireshark and 47 % in Samba. However, the individual firm contribution ranges from low to very high. Table 4 presents the number of messages and centrality degree of top 10 active firms in mailing list. In Wireshark project, the maximum value of centrality degree of Philips is 48, meaning that they are in contact with 48 other firms. In Samba project, the maximum value of centrality degree of Redhat is 71, showing that they are in contact with 71 other firms. The top three firms account for 60% and 56% of the mails in Wireshark and Samba, respectively. We interviewed two firms in these lists (D1 and D5) for answering RQ2 (Section 5).

**Table 4: Number of entries and centrality degree in mailing list**

| Wireshark | | | Samba | | |
|---|---|---|---|---|---|
| **Firm** | **Entries** | **Degree** | **Firm** | **Entries** | **Degree** |
| **Philips** | 1195 | **48** | **Redhat** | 4480 | 71 |
| **Ericsson (D1)** | 1322 | **39** | **Sernet** | 3765 | 66 |
| **AT&T** | 756 | **34** | **Google** | 1835 | 57 |
| **Trihedral** | 222 | **21** | **IBM (D5)** | 1701 | 48 |
| **Thales** | 548 | **19** | **HP** | 1408 | 44 |
| **Mxtelecom** | 149 | **19** | **Eurocoopter** | 874 | 35 |
| **Gtech** | 165 | **13** | **SGI** | 335 | 29 |
| **Detica** | 64 | **10** | **Padl** | 82 | 29 |
| **Csr** | 67 | **10** | **Zylog** | 159 | 28 |
| **Sequans** | 31 | **10** | **Nokia** | 104 | 28 |

Figure 6 presents the map of our interviewed cases in the social structure of OSS projects. The selection process ensured that interviewees participated in the projects for a sufficient duration. We can see that the interviewees come from different layer of the projects, hence, representing for the whole projects.

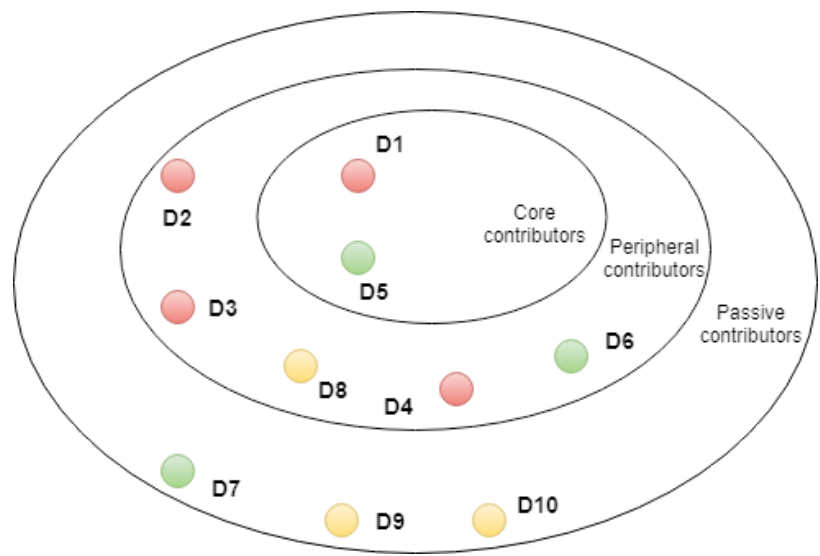

**Figure 6: Social positions of interviewees in OSS projects**

# 5. RQ2: How do commercial firms manage coopetition with other firms in such context?

By investigating communication patterns among firms in OSS projects and analyzing interview transcripts via the thematic analysis, we proposed a Coopetition in Open Source Software (COSS). The model is grounded from thematic concepts that extends our research presented in Section 2.3. The COSS captures the underlying phenomenon of firm participation in OSS projects from coopetition perspective. The main concepts representing the underlying phenomenon have been grouped together to form high level categories, as seen in Figure 7. The model is centralized around the concept of Coopetition. Beyond the concept of coopetition in business research that consists of Competition and Collaboration, we identify two additional dimensions of the concept, which are Gatekeeping and Firm awareness. Coopetition activities are visible with the recognition of firm boundary in the projects and implemented via gatekeeping mechanisms, which are synchronizing code, strategic filtering and navigating information flow. Antecedent factors that influents Coopetition concepts include structural condition, trust, perceived benefit, and strategic vision. Structural condition includes two sub concepts, public communication and direct communication. Consequent factors of coopetition include organizational learning, knowledge sharing and task effectiveness. Following subsections below describe the grounded evidence for each model's elements.

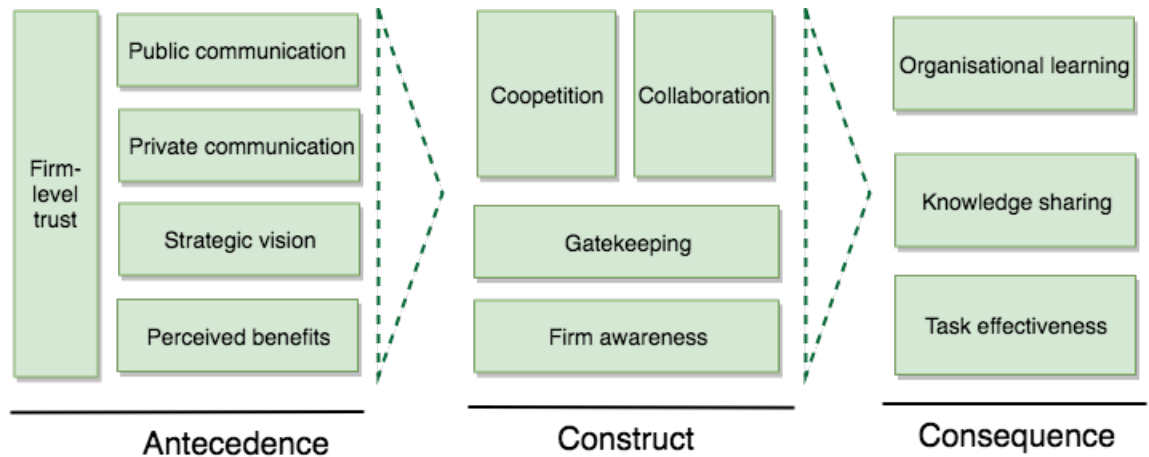

**Figure 7: The model of Coopetition in Open Source Software (COSS)**

## 5.1. Public communication

The public communication channels used in our OSS projects were the mailing list and bug tracking systems. In both projects, the distribution of public communication is highly right-skewed, as shown in Figure 8. In Samba project, Sernet has contributed almost 35% of total number of message via mailing list. The top three firms account for 60% and 56% of the mails in Wireshark and Samba, respectively.

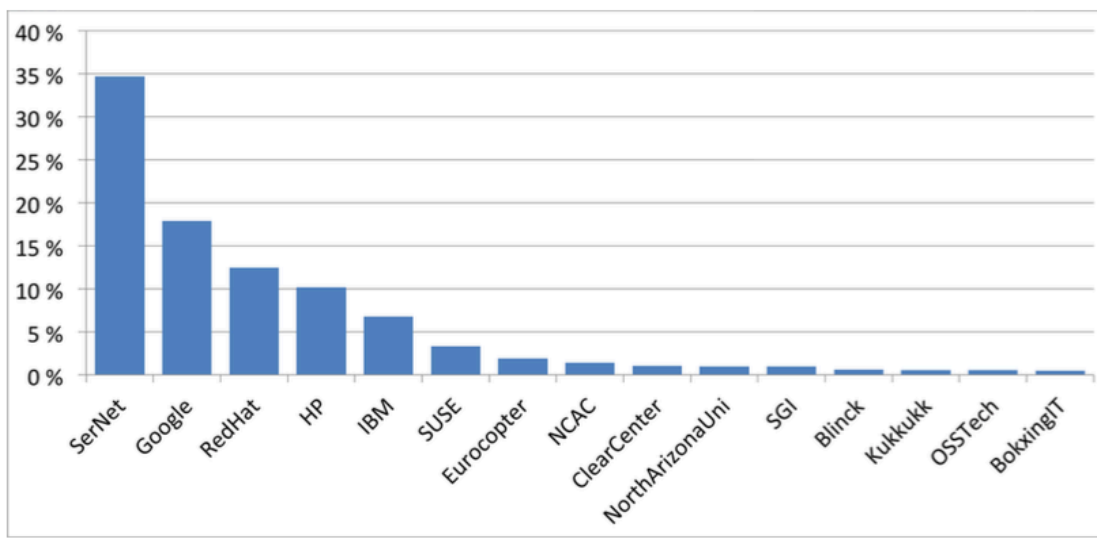

**Figure 8: Distribution of number of mails per firm in Samba**

Developers mentioned several incentives for using such channels, for instance, they use the public channels for discussing, participating and/or influencing the ongoing development. D4 mentioned that he publicly asked questions, discussed ideas and found collaboration via public channels: "*Basically, the times when I need guidance or I have a problem, or answering other people's questions, whether it is other developers or users or whatever. Or if I have an idea about something. (...) I made a suggestion 'hey maybe we should do something to catch this problem automatically in the build-bots rather than ...' Anyway, just making suggestions and putting them out basically*." D6 considered mailing lists as a traceable information storage that is useful for his job: "*Usually all discussions are done on the mailing list (...) this way we have a history of all discussions. I participate in discussion either to help someone with Samba or to make my point in area of my interest at the moment.*" Influencing project features by participation is one incentive expressed by D1: "*If they are working on something that I see as usable for us internally, we find it interesting. It is smart to participate in the discussions when they are doing the development, and not come in afterwards. That is because while they are doing the changes and the development, they are more open for suggestions for changes and improvements.*"

Asking for guidance and support on mailing lists is common, however some developers underlined that they did not ask for solutions to their problems here. Rather, they would ask for useful advices and a push in the right direction. D3 stated that "*Sometimes I have sent emails to the development list and said that I am confused by this, can someone shed some light on it.*" Developer D4 expressed a similar approach in: "*More often I will ask people 'OK, I have this problem and I am trying to solve it. I can see two ways to solve it, does anybody have an opinion on which way is the better way?*" By this way, technical issues within a firm can be discussed and supported by external people.

D2, D3 and D4 said that they asked questions about architectural decisions in the public channels. Posting features requests or interesting ideas is also common, and some of the interviewed developers find it motivating to describe their ideas and approach to the other community members. By this way the feature expectation is communicated and other developers can come with suggestions and even join the development. D5 and D6 stated: "*I tend to participate in discussions where I feel I have a useful technical contribution to make.*" (D5) *and* "*I participate in discussion either to help someone with Samba or to make my point in area of my interest at the moment.*" (D6)

## 5.2. Private communication

Firms use private communication for many purposes, including both cooperative and competitive manners. Developers mentioned that they had used direct and/or private communication channels for asking for help from the domain experts in the project. Communication channels used are e-mail and instant messaging, Skype and telephone. D3 said: "*I have done it [contacted developers directly] some times in the past. Not just as a general I am stuck, can you help, but because it would be an area I knew the other guy was working on.*" The private communication is usually the result from a gradual establishment via public communication, as mentioned by D6: "*Usually I tend to do R&D tasks myself. I often seek for reviews of my work. When I need the assistance, I will go directly to a developer in the community.*"

Comparing to public channels, D8 considered private communication as a way to establish high-quality contact points and potential collaboration for further projects. He mentioned that a fork from project mainstream should probably include best developers in the community who are not necessarily the guy in the "*onion core*". It is also stated that a private channel is a quick and efficient communication medium. D9 explained that he used instant messaging for contacting developers in the community when he wanted a quick feedback. Private communication seems to be in favor comparing to public communication. D9 mentioned: "*We try to address as much as we can the issues that come to us… Normally if we get a private message about an issue, we will take it with higher priority …*" D5 mentioned that when discussing legal or security sensitive issues, he used a private communication channel. The nature of such issues invokes the use of private channels as posting it in the public channels may result in security breaches or similarly bad situations. Although none of the other developers said anything about the use of direct channels for such issues, we believe that it is a common procedure in most OSS projects.

## 5.3. Trust

Trust is one of the fundamental traits of a successful collaborative environment [49,67-69]. Raymond stated that "*open-source culture has an elaborate set of customs…[which] regulate who can modify software, the circumstances under which it can be modified, and (especially) who has the right to redistribute modified versions back to the community*" [70]. In our cases, interviewee stated that the success of OSS projects is meaningful to them. For instance, with the advance of the Wireshark tool, D4 can use it to serve for his daily work. Based on trust, developers can collaborate for the sake of their OSS project. D3 said that they have contacted trustable developers directly to avoid asking silly or dumb questions in public: "*I got relationships with other developers and sometimes we don't want to ask in mailing list causes it is a really stupid question and you do not want to ask the whole mailing list, so you just ask the guy you trust*". When a developer needs help to design a code or fix a bug, other developers would be willing to assist. By helping one another, developers demonstrate their skills and knowledge, which develops a positive expectation of competence and reliability. Level of trust is related to the status of the developers in OSS projects, which is evident in the following section. The observation is aligned with previous research on the role of trust in successful interpersonal relationships [37,38].

## 5.4. Perceived benefit

Despite the risks associated with competitors, many firms decided to be open in sharing and synchronizing their source code with OSS communities. Source code can be synchronized with upstream development in OSS projects, for instance, described by D5: "*In general, our philosophy is to develop upstream first and then back-port changes that have been approved by the upstream community into our products. We stay very involved in the communities and try to keep the differences between our packaged software and upstream software to the minimum necessary.*" Firms perceive benefits with such involvement as **avoiding maintenance and merging issues** when combining public parts of private parts of source codes. D10 illustrated for this idea: "*… if you are to make a change in the core, and you want to keep it private, you will have to fork the project and maintain it yourself. (...) I believe, in the general case, that you gain more from contributing to the development, that retaining your code from the community*". D1 mentioned that "*We do not have to maintain our own code base and synchronize it. We just commit code to the source and have it there. If we had not had the* *commit access as easy as I do, we could have had our own version of Wireshark and the sources, but then we would have to do more work in merging our version with the new releases of Wireshark.*"

Firms also concern about their social positions in the projects. It is apparent that a central position in the community is closely related to being a core developer in most cases. Two benefits mentioned by the interviewees are: (1) easier code inclusion and thereby avoid the need of having a private code repository, and (2) receiving more help from other community members. D4 highlights the importance of social position in OSS community: "*I think it [having a position] helps a lot. I think there is a difference if, lets say, D2 asks for help, then I will help him if I can. But if [Developer Name] from I have never really heard of, is asking for help then my level of effort is usually lower. And part of that is because I know D2 personally, and part of that is because I know that he does a tremendous amount of work. My view is that if he needs help he deserves the help. And I think it goes the other way too, if people are more likely to help me because of the contributions I have made and they know that I have been contributing for a long time. I think it helps to have some sort of status within the community.*"

## 5.5. Strategic vision

The role of strategic vision on firm participation is somewhat vague in our cases. Firm's strategy could be how a firm develops and deploy their product, i.e. how external resources are used to reduce development and maintenance cost. The vision of firm's strategy needs to be aligned at not only managerial but also operational levels. The transfer of strategic visions is not clearly evidenced in our cases. For instance, a developer D4 mentioned he spent significant office work hours as well as spare time on contributing to Wireshark. He acknowledged the benefits other developers in his firm received from his participation in the OSS project and the fact that he freely participated in Wireshark: "*it is not an official part of my job, but a lot of the developers, testers and the customer support people use Wireshark extensively.*" However, his firm lacked formal strategies to decide how developers shall participate and develop the OSS, what code that shall be contributed back to official sources, and how to maintain the OSS knowledge base within in the firm.

## 5.6. Gatekeeping

The perceptions of a gatekeeper, who navigates information flows between his/ her firm and external actors, were acknowledged by all interviewees. While firms might have different needs and work practices, gatekeepers are the ones stay in between the firm and the OSS project in some way, as shown in Figure 9. D1 stated that when his coworkers found issues with the third party components, they informed D1, but not project managers. D7 expressed a similar perception: "*Yes, I act as a bridge between [Firm Name] and Samba and forward bugs/errors to the community.*" The gatekeeper is often an active actor in contributing to the community, as mentioned by D2: "*Many of our core developers are working for smaller companies, and have a responsibility for the internal protocols that their company needs. (...) I think most developers work individually, and have the role of providing Wireshark functionality to the other developers in the firm.*"

In a cooperative manner, the gatekeeper is the hub of information and issues that can be reached by different developers across the organizations, as stated by D4:"*Yes, everybody definitely knows that I am the Wireshark guy. All the developers, testers and customer support people know that they can come to me if they have Wireshark issues...*". In firms with multiple developers active in upstream development, i.e. committing to OSS projects, there is often a recognized gatekeeper role among them. D5 mentioned: "*In general when it comes to contributing patches upstream each developer in [Company Name] is independent and can directly approach the upstream project… The [Company Name] Samba package maintainer usually has a task of being the gatekeeper for those bugs that have been reported against [Company Name] products by the customers or the support teams...*" In this case, while code is contributed independently by individuals in the firm, the bugs is managed by a gatekeeper who submits bug reports on behalf of the firm into the OSS project's bug tracking system.

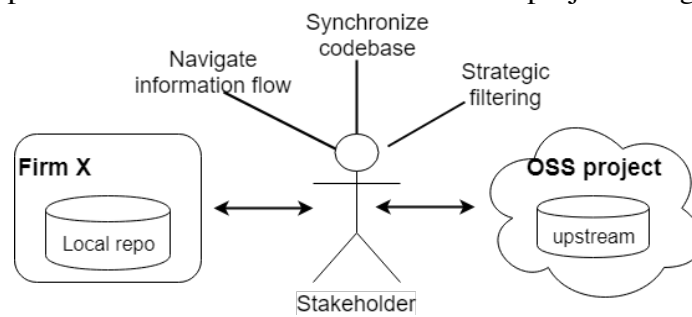

**Figure 9: The role of gatekeeper in a commercial firm**

In a competitive manner, gatekeepers would make sure that not all private source code be revealed to public. Firms might contribute code that relate to core components of OSS products, or utility functions. In a typical scenario, firms maintain their private repositories, where many components are parts of firms' core values. Such components should not be revealed, as mentioned by D4: "*The majority of the stuff I have written for Wireshark has been pushed up… But you sort of draw a line in the stuff that is obscure enough to not push. The only people who should be looking at our proprietary protocol should be us…*" Some of the code is regarded as proprietary and is retained in the firm's private code repository, due to technical specific, or legal and authorization issues D2 mentioned: "*Mainly protocol dissectors for protocols used in our equipment, if the protocol is based on open protocol descriptions from 3GPP, ITU or IETF (RFC) it is considered OK to make an individual contribution to OSS...*" Code which is not relevant, sensitive or poorly written would be filtered out by gatekeepers, as mentioned by D4: "*The stuff we do not send in is stuff that is not of interest to anybody except us. (...) And the other part is that I do not think the company would be thrilled by a publication of these protocols. In order to push those things to Wireshark I would need to get authorization.*"

## 5.7. Firm awareness

Several interviewees acknowledged the presence of at least another firm in the community (D1, D2, D3, D8, D9, D10). However, developers remark that it is not the knowledge of what other firms work for that is valuable, rather it is the knowledge of what business domain they are working within. D2 replied when was asked about other firm awareness: "*Yes, but I do not* *know that much about the firms of the other developers. They typically say that they work for Firm X, and that is it. What firm they are working for is not that important to me.*" D3 emphasized the potential value of having the firm awareness: "*... I know that D2 may have some role as a contact for Firm X... I know that D2 may be someone who is good at getting log files for specific things. In the past when I was working with voice over IP, I thought sometimes he was able to give me some log files from within his company, but I did not really think of him as the company representative. I think of him as a company person who may be able to get logs for me, like he does.*" In Bootstrap, developers expressed the concern on how other firms were doing related to the web technology, in order to draw lessons learnt for their product vision. D8 mentioned: "*We care about if other company are using this technology in their products, so we can learn from them... We do not care if some guys just want to play with the technology ...*"

Additionally, the interviewees were asked if they considered that their contributions could be used by other firms to gain competitive advantage. The majority dismissed this perception, for example: "*As Firm X does not directly control Wireshark, I guess we have to be a bit careful when we are in contact with other developers. (...) I believe, in the general case, that you gain more from contributing to the development, that retaining your code from the community*", stated by D2. A final remark by D5 about the competitiveness is: "*Although there may be some competition between companies, as engineers we seek collaboration for mutual benefit. We already know any advancement will be used by everybody, that is not a problem, we get back as much as we give out.*"

## 5.8. Collaboration

Although collaboration within an OSS community is typically informal and not planned, there are matters that have to be decided upon. For instance, when there is a new post in a mailing list, a developer has to decide whether to engage in the discussion with the others or not (essentially collaborating with them). The awareness of other firms in this aspect may prosper the collaboration. Firm- paid developers with similar needs and interests can collaborate and draw on each other's abilities. Knowing that a developer works for a certain firm, and that he can provide certain code artifacts also influences the collaboration. Establishing relationships to such valuable developers through collaboration is key. There is a strong desire to return favors and honor developer's positions by assisting them when they need help.

Many commercial firms adopt OSS, but do not participate nor contribute back to the OSS communities. Some of these firms collaborate directly with others to develop OSS-based products further, with or without participating in the OSS community. How to perform the collaboration is an aspect firms have to decide. As described above, the collaboration can take place within the OSS community using public or private communication channels, or outside the community using private channels and private code repositories.

## 5.9. Awareness of competition

Firms working within the same business domain are often competitors in the market, and thus it is interesting to see how influential the firm awareness is when firms come together in community based OSS projects to develop software collectively. Surprisingly, firm-paid developers said that they perceived other developers as partners and/or friends rather than competitors. D5 pointed out that he had met many developers at the OSS developer conference, and considered many of them as friends. D1 explained that he did not make any distinction between a firm-paid developer and a volunteering developer: "*I think of them as developers, and not about which firms they represent.*" D7 said that he would perceive others as partners. D6 mentioned: "*I have always thought of others as partners. Even more - I think about them as colleagues.*" D4, D8 and D9 shared similar thoughts, and dismissed the perception of other firm-paid developers as competitors: "*I guess as things have evolved we do actually compete in some aspects with some of these people at this point. But that hasn't really occurred to me much… I have noticed more people who tend to be customers of ours, rather than true competitors. We might be competitors within some areas, but I have never really thought about it I guess*", stated by D9.

The issue of competition from a firm from somewhere else in the world might not be significant for a startup and a SME who focus on having their product released as fast as possible. Without a clear vision on how their market or technical advantages are influenced by sharing and using OSS source code, the concern of competition is not much relevant. D8 also mentioned: "*…you think about other firms as your competitors, but I do not think that really comes in to my interactions really. They have their own users somewhere around the world…. I have sometimes seen contributions from their developers, but I think that is good…*" Consequently, the coopetition concept in these OSS projects might be very much cooperation-dominant.

Another observation is that the firm's social position is not used by any firms to dominate OSS development. D6 mentioned: "*Before working on Samba I used to think that big companies may have big influence in OSS project simply by "buying" core developers. Now, that I know most of the people working on Samba, I know that this is not feasible.*" Hence, having a position, or "*buying*" one, is not the way firms relate to nor influence the OSS development.

## 5.10. Consequent factors

Interviewees acknowledged the benefits of participating in OSS projects, including knowledge sharing, organizational learning and task effectiveness. D2 mentioned that many best practices found in reviewing code and proper comments on commits. He also appreciated the activeness level of the project with fast feedbacks. The practices are acknowledged and brought into consideration for improvement at his team. Maintaining an awareness of the other developers and what they are currently working on is also recognized and is promoted by D6 in his firms for avoiding duplicated code across the whole codebase. Organizational learning also occurs at the project level. When a firm observes the participation and interaction of core firms in the OSS projects, they can infer strategic focus areas from, i.e. feature requests and application cases.

In our cases, in-house product development depends on the OSS projects by (1) using tools as outcomes of the projects or (2) integrating and building their products on top OSS components. The dependence infers that a task that relates to OSS codes is collective performed and the task scope is beyond the OSS project. In a cooperative-dominated environment, the task will be done in an easier way. In a competitive-dominated environment, the awareness of competitors might be harmful for jointly completing the task. However, this is not directly evident from our cases.

# 6. Discussions

Table 6 summarizes our findings in the comparison with existing literature. While many findings confirm existing knowledge, they also provide some novel findings. This section will discuss our findings based on four topics: centralized communication structure in community-lead OSS projects (Section 6.1), modelling coopetition in the context of OSS projects (Section 6.2), the role of a gatekeeper in implementing coopetition strategies (Section 6.3) and firm contribution strategy in OSS projects (Section 6.4). Each section will discuss our findings with related work. The final section presents our actions to address threats to validities (Section 6.5).

## Table 5: Summary of findings

| Findings | Type | Current knowledge |
|---|---|---|
| OSS infrastructure as foundation for both public and private communication among firms | Confirmation | structures as those in OSS enables the integration of external resources [56] |
| Firms activities are visible in OSS projects | Confirmation | heterogeneity exists between firm-paid developers and voluntary developers [25, 58] |
| Some firms in the core positions, most of firms contribute little | New | Onion-like structure at developers level [47,48] |
| Coopetition exists among firms | Confirmation | strong explicit governance approaches can directly affect other firm's benefits [64] |
| Cooperation-dominated coopetition among firms at code and issue levels | Confirmation | competition for the same revenue model does not necessary affect collaboration within OSS projects [10,11, 29, 62] |
| Gatekeepers provide a mechanism to perform coopetition | Contradict | developers within a firm need to be divided to take charge of either collaboration or competition [9] |
| Trust is the foundation of establishing communication, collaboration and also competition | Confirmation | Trust as a success factors in collaboration in OSS projects [38,39] |
| Strategic vision is not significant at developers' level | New | Sharing strategic vision is also critical for collaboration at team level |
| Firms gain social position in OSS projects, avoid merging and bug fixes, impact on influencing development and | New | Perceived benefit is associated with both cooperative and competitive attitudes [35,41,42] |

get supported

## 6.1. Centralized communication structure in community-lead OSS projects

Commercial firms participating in community-based OSS projects collaborate in various ways across the organizational boundaries. Crowston et al. stated that communications structure of a project is an important element in understanding a project's practices [48]. In our cases, the majority of the activity in OSS projects is generated by a small subset of the firms, and that the remaining firms participate with little to none contribution. Wireshark and Samba demonstrate a communications centralization structure as in the onion-like social structure model [48]. Oezbek et al. investigated eleven OSS projects and revealed that the role of a developer in the core layer might be more important than the fact that they do (commit code, fix bug, answer emails, etc) more [76]. Our quantitative analysis of Wireshark and Samba confirmed these results by showing the dominant contributions of developers and firms in the core layer. Our qualitative data revealed possible importance of these developers in implementing firms' strategies, i.e. collaboration or competition. Dewan et al. showed that the heterogeneity, which exists between firm-paid developers and voluntary developers shapes the evolution of OSS community and its product [58]. In our case, we showed that even firm-paid developers have significant contributions to code commits and communication, it is not significantly different between firm-paid and voluntary developers. From communication structure, this reveals a different finding from Dahlander's work [25].

## 6.2. Modelling coopetition in the context of OSS projects

Business literature mentions the difficulty of identifying coopetition in a real world context [26]. Dagnino et al. highlighted that coopetition does not simply emerge from joining competition and collaboration, but they mix together to form a new kind of strategic interdependence between firms [26]. We agree and illustrate for this view by showing that in OSS projects, commercial firms focus on activities that create a common value with an awareness of not sharing their technical and legal sensitive information. From our cases, COSS validates at the meso level of strategic collaboration, where firms within the same or similar domain collaborate. Among antecedent factors from literature, we highlight the role of a structural condition via public and private communication infrastructures. The transparent and effective communication infrastructure provides a mechanism for coopetition. Our study describes a competition-dominated type of coopetition. Even when firms are aware of their competitors, the attitude of collaboration is still overwhelming. Valenca et al. raise a question whether firms are collaborators or competitors in software ecosystems [14]. At the requirement engineering level, the authors found several significant challenges among firms within the same collaborative network [14]. OSS projects and firms might have divergent interests but firms can manage to discover areas of convergent interest and be able to adapt their organizing practices to collaborate [3]. In our case, this is clearly observable at the operational level. The finding also matches with observations by Linåker [10].

## 6.3. The role of a gatekeeper in implementing coopetition strategies

Bengtsson et al. argued that individuals within a firm could only act in accordance with one of the two logics of interaction at a time, i.e., either to compete or to collaborate [9]. Our observation on a gatekeeper role gives an alternative explanation on how firms manage such scenario. The firm's strategy can be flexible, for example fully open core sourcing at one time, and filtering of shared code at another time. The implementation of such strategies is done via the firm gatekeeper, who does actual technical contribution to the community. Therefore, in contrast with Bengtsson's findings, we find that it is possible to implement a firm-level dynamic interaction via individuals in software projects. The role of a gatekeeper is discussed in the context of commercial distributed software teams [72,73]. Marczak et al. found the role of knowledge brokers who would have a significant impact on information flow in requirement-interdependent teams [73]. In a context of firm-to-firm interaction, we showed that a gatekeeper could navigate the information flow beyond firm's boundaries. Nguyen-Duc et al. showed four common tasks of a gatekeeper: task negotiation, conflict resolution, task- related information navigation and boundary object setups [72]. While the authors investigated gatekeepers in a software firm and a OSS project separately, this work focuses on boundary spanning activities between the OSS communities and software firms. By influencing the gatekeepers, managing code flows and information flows, firms can implement competing or collaborating strategies.

## 6.4. Firm contribution strategy in OSS projects

There exist some studies capturing the phenomenon of commercial firms contributing to OSS projects. Linåker et al. investigated contribution strategies of firms when participating in OSS projects [74]. The authors proposed the Contribution Acceptance Process (CAP) model to determine if source code or any types of contributions can be contributed or not. The CAP model bases on two dimensions: (1) the benefits company can receive and (2) the knowledge behind the contributions to acquire and control [74]. While these two dimensions are similar to our model's elements: perceived benefits (Section 5.4) and gatekeeping (Section 5.6), our model also explore other factors that impact the ways firms contribute to the OSS communities and collaborate with other firms. Munir et al. discussed how the openness of software firms might help them to gain benefits from OSS communities from four dimensions: (1) strategy, (2) triggers, (3) outcomes, and (4) level of openness. The model is similar with some elements in our COSS model, i.e. strategic vision, communication, gatekeeping and consequent factors. However, these models do not capture the competition strategy that firms might adopt in OSS projects. Unlike the previous work, our COSS model proposed a comprehensive view on factors that impact the strategy of collaboration and competition.

## 6.5. Threats to validity

### 6.5.1. Construct validity

Threats to construct validity consider the relationship between theory and observation, in case the measured variables do not provide a good measure of the actual factors [66]. In a qualitative study, construct validity can be thought of a "labeling" issue, as we might find the construct of the outcomes that we believe we are trying to capture. A main assumption in our study lies in the way we identify coopetition among commercial firms. As the coopetition concept comes from economic and business research, we did not have a direct map from how the concept operationalize in SE research. Previous studies that mention term "coopetition" [10, 14], do not provide the construct of this concept. Hence, to our best knowledge, this is the first study in SE attempt to operationalize this concept. We reduced this risk by a detail review and the identification of characteristics of coopetition, the exploration of the context where the construct is investigated. Both quantitative and qualitative data was collected in concept's elements and summarized in the end to describe the model. We also include discussion with co-authors and an expert in the entrepreneurship in validate our observation.

The phenomenon is operationalized based on public and private communication among developers participated in OSS projects. We were aware of other communication channels, such as private messaging, telephone and Skype, however, we do not have a feasible way to quantify this. We limited the investigation in public collaboration where developers responsed to the same mailing list or comment on the same issue. Regarding to the identification of firm participation, we used SNA with density metrics, such as degree centrality and closeness [65]. Other network-based measures for the same construct (e.g., transitivity, compactness, and connectedness) could be considered for enhancing the rigor of this research. We also used an unweight approach to perform SNA, which ignored the firms' characteristics, such as firm size, and business strategy towards the OSS community. This could be considered in future work, especially in firm-based OSS projects.

The risk of operationalization is reduced by using a mixed method research, including both quantitative and qualitative data. The interviewees were conducted with firms from different social position in OSS projects, which increase the credibility in the observation of phenomenon. The data is limited at ten interviews. However, we had reached data saturation [66] when interviewing Bootstrap case. Although, interviewees were selected from different types of OSS projects, different company profiles, we found that their responses were consistent, which increase our confidence in the trustworthiness of the data.

### 6.5.2. External validity

This threat considers the ability to generalize our findings. The goal of this study is not to achieve statistical generalization, but rather an analytical generalization. This is particularly important when studying a complex phenomenon, in our case is coopetition in OSS projects. To avoid the bias on findings from a single case, we analyzed two OSS projects. Qualitative data was further collected from the third OSS project to improve the generalization. With the in-depth investigation in both community and firms' sides of the projects, we are confident about the explanation power of the COSS model for similar contexts. Our OSS projects produce a library, a framework and an application, employing GPL and MIT licenses. Our cases represent for a community-initiated OSS projects, that are initiated and lead by the community.

Further research should replicate our method on other types of OSS projects to explore other collaboration and competition scenarios. They are also popular OSS projects with years of operation, hence the products and collaboration process have been stable. The findings might not be directly applicable to emerging OSS projects, or projects initiated by firms. Research on projects with different types of OSS licenses might lead to a variety in our model.

### 6.5.3. Reliability

This threat concerns about the level to which the operational aspects of the study, such as data collection and analysis procedures, are repeatable with the same results. The main data collection was done as a part of a master thesis. All interviews were recorded and transcribed verbatim in order to make sure that no data reduction occurred prematurely. The transcription of the interviews was reviewed and interpreted by the other author. In case of vague statements, one author is responsible for follow-up discussions with interviewees for clarification. We used both quantitative data about communication among firms in the project and qualitative data from interviews of firms from different contribution layers. The data triangulation allows our findings represent the true situation of investigated projects. Moreover, the paper has gone through proof-read from several senior researchers in the domain. Their feedbacks help us to improve the paper significantly since the first draft.

## 7. Conclusions

Coopetition is an important topic in economics and business research [9, 26], but it is overlooked in other domains. In modern software industry, the popularity of developing software products beyond firm's boundary makes coopetition a relevant theme. In this paper, we used both qualitative and quantitative data to investigate coopetition in OSS projects. Firstly, we found that commercial firms participating in community-initiated OSS projects collaborate in various ways across the organizational boundaries. While most of firms contribute little, a small number of firms are very active and account for large proportions of contribution. It is also evident that firms interact across their boundaries in OSS projects. Secondly, we proposed an empirical model COSS to explain for root causes of coopetition in OSS projects. The COSS model shows that coopetition is based on the firm awareness, structural condition of the OSS projects and operated by gatekeepers. The coopetition is cooperation-dominated even among firms working in the same business domain with similar business models.

The findings have implications for research. We offer a descriptive explanation of how coopetition occurs and impacts in OSS projects. We observe that software firms emphasize the co-creation of common value and partly react to the potential competitiveness in OSS projects. The highlight of our findings is the COSS model, which argues that competition and collaboration can both be handled by gatekeepers. The role of gatekeepers in crossing organizational boundaries is still an interesting research topic. For SE with abundant research on OSS collaboration and communication, the study on inter-firm coopetition is a novel way of looking at the same data sources and infrastructures.

The study also has implications for practitioners. We offer software firms insights about different coopetition strategies observed in a community-driven OSS project. For instance, not all communication goes through the public channels in OSS projects. Legal and security sensitive issues commonly go through private or closed channels because of their natures. Furthermore, firms should consider a gatekeeper as an important role when they plan to participate and gain benefit from OSS projects.

For future work, the next step would be to validate the COSS model with a larger set of cases. Our research here only uses three community-driven OSS projects, which limits the generalization of findings. Moreover, a longitudinal observation on how coopetition evolves among firms can provides knowledge that goes beyond cross-sectional observations. Last but not least, further investigation about employing the role of gatekeepers for coopetition is needed to provide actionable guideline for successful operation of inter-firm coopetition. Future work can also investigate OSS project settings that affect firm collaboration, i.e. OSS license, and feature request mechanism. It would be interested to see how these factors could play a role in our model.